\documentclass[aps,prapplied,reprint,superscriptaddress,raggedbottom]{revtex4-2}

\usepackage[T1]{fontenc}
\usepackage[utf8]{inputenc}
\ifdefined\pdfgentounicode
  \input glyphtounicode
\fi
\usepackage{amsmath,amssymb}
\usepackage{graphicx}
\usepackage{bm}
\usepackage[hidelinks]{hyperref}
\newcommand{\UnitMicro}{\text{\textnormal{\textmu}}}

\newcommand{\NumUnit}[2]{\ensuremath{#1\,\mathrm{#2}}}
\makeatletter
\AtBeginDocument{%
  \renewcommand{\bibsection}{%
    \par
    \addvspace{12\p@}%
    \begingroup
      \baselineskip26\p@
      \bib@device{\columnwidth}{0.3\columnwidth}%
    \endgroup
    \nobreak\@nobreaktrue
    \addvspace{19\p@}%
    \par
  }%
}
\makeatother
\newcommand{\FigureOneWidth}{0.70\textwidth}
\newcommand{\FigureTwoWidth}{0.80\textwidth}
\newcommand{\FigureThreeWidth}{0.90\textwidth}
\newcommand{\FigureFourWidth}{1.00\textwidth}
\newcommand{\FigureFiveWidth}{0.88\textwidth}
\newcommand{\FigureSixWidth}{1.00\textwidth}
\newcommand{\FigureSevenWidth}{0.90\columnwidth}
\begin{document}
\frenchspacing
\tolerance=1000
\emergencystretch=1em
\title{Characterization and Suppression of Parasitic Back-Reflection Noise in Transponder-Type Heterodyne Interferometry}
\newcommand{\InstitutionOne}{State Key Laboratory of Precision Measurement Technology and Instruments, Department of Precision Instrument, Tsinghua University, Beijing 100084, China}
\newcommand{\InstitutionTwo}{Key Lab of Luminescence and Optical Information, School of Physical Science and Engineering, Beijing Jiaotong University, Beijing 100044, China}
\author{Xiaoyang~Lin}
\affiliation{\InstitutionOne}
\author{Xin~Xu}
\email[Corresponding author: ]{xinxu1@bjtu.edu.cn}
\affiliation{\InstitutionTwo}
\author{Xinhang~Lyu}
\affiliation{\InstitutionOne}
\author{Peijia~Tian}
\affiliation{\InstitutionOne}
\author{Yujie~Feng}
\affiliation{\InstitutionOne}
\author{Mingkuo~Hai}
\affiliation{\InstitutionOne}
\author{Yidong~Tan}
\affiliation{\InstitutionOne}
\author{Yan~Li}
\affiliation{\InstitutionOne}
\author{Haoyun~Wei}
\email[Corresponding author: ]{luckiwei@mail.tsinghua.edu.cn}
\affiliation{\InstitutionOne}

\begin{abstract}
Transponder-type heterodyne interferometry is a versatile metrological technique for high-precision measurements, with applications in Earth's gravity-field recovery, space-based gravitational-wave detection, deep-space ranging, and Doppler tracking. However, parasitic back-reflections can contaminate phase readout and degrade measurement sensitivity. To characterize and suppress this noise, we constructed a transponder-type interferometer with coaxially aligned counter-propagating beams, achieving stable operation at a received optical power of \NumUnit{720}{pW}. An adjustable reflecting system was introduced into the master-side transmit path to simulate back-reflection under realistic conditions. Optical path scans and model fitting showed that phase errors vary periodically with the relative phase between the main and parasitic interference signals, which defines the working point. Assisted by \(\pi\) measurement, adjustment to low-sensitivity working points was combined with balanced detection to suppress back-reflection phase noise, achieving maximum suppression factors of 77 for single-frequency phase errors and 23 for band-limited phase-noise amplitude spectral density. At a back-reflection power of \NumUnit{1.5}{nW}, \(\pi\) measurement remained effective for working point diagnosis. Its quadrature counterpart, the beat-amplitude difference, can distinguish opposite path offsets near a low-sensitivity working point. These results provide an effective approach for in situ back-reflection characterization and phase-noise suppression in transponder-type heterodyne interferometry.
\end{abstract}
\maketitle

\section{Introduction}\label{sec:1}

High-precision laser interferometry is needed for Earth's gravity-field measurement \cite{ref1}, space-based gravitational-wave detection \cite{ref2,ref3,ref4}, deep-space ranging, and Doppler tracking \cite{ref5,ref6}. In inter-satellite and deep-space optical links, propagation losses limit the received optical power. For example, space-based gravitational-wave interferometers (SGWIs) operate with sub-nanowatt received optical power \cite{ref7}. Returning such a weak beam by passive reflection would incur further propagation losses, increasing the shot-noise contribution to the phase readout. These systems therefore typically employ transponder-type heterodyne interferometry \cite{ref1,ref6,ref7,ref8,ref9,ref10}, in which the slave laser is phase-locked to the weak incoming beam and transmits a high-power beam back to the master.

In a transponder-type interferometer, the weak received beam coexists with a much stronger transmitted beam, where back-reflection occurs when a fraction of the transmitted light returns toward the optical bench. Back-reflection can originate from optical elements \cite{ref11,ref12}, backlink fibers \cite{ref13,ref14}, and telescope mirrors \cite{ref15,ref16,ref17,ref18}, among which the telescope system is of particular concern because it contains multiple reflective surfaces, carries high transmitted optical power, and is mechanically connected to the optical bench through supporting structures \cite{ref15,ref18}. The relative phase between the main and parasitic interference signals defines the working point. The coupling of back-reflection optical-path fluctuations into the measured phase depends on both the back-reflected field amplitude and the working point \cite{ref19}. At a low-sensitivity working point, small optical-path fluctuations produce a reduced phase response.

Studies of back-reflection have addressed both its optical origins and its impact on interferometric phase readout. At the optical-system level, stray light modeling has been used to identify parasitic paths \cite{ref17,ref20}, while optical-path optimization \cite{ref12,ref21}, baffle design \cite{ref22}, and surface treatments \cite{ref23} aim to reduce the light returning to the interferometer. Residual back-reflection nevertheless requires further mitigation at the readout stage. Balanced detection combines the signals from the two output ports of a recombination beam splitter to suppress certain stray light contributions, although imperfect balance can leave residual coupling \cite{ref13,ref15,ref24,ref25}. \(\pi\) measurement, formed from the phase difference between these ports, has been used as an auxiliary diagnostic of parasitic interference \cite{ref24,ref26}. Optical path control has been used to suppress optical path noise and parasitic coupling in heterodyne interferometers \cite{ref27,ref28,ref29}. In particular, Cao et al. demonstrated noise reduction by locking the stray light optical path length at a zero-coupling point \cite{ref29}. These studies motivate combining \(\pi\)-assisted working point adjustment with balanced detection to suppress residual back-reflection phase noise in transponder-type heterodyne interferometry. The amplitude method has been used to evaluate periodic errors in heterodyne interferometers \cite{ref30,ref31}. For weak back-reflection, beat-amplitude modulation is in quadrature with the phase error to first order \cite{ref30}, suggesting that it could serve as an auxiliary signal for monitoring a low-sensitivity working point.

In this work, we characterize parasitic back-reflection coupling and develop a noise suppression method combining \(\pi\)-assisted working point adjustment with balanced detection. Experiments are performed in a transponder-type interferometer with coaxially aligned counter-propagating beams, where an adjustable reflecting system simulates back-reflection under realistic conditions. Section~\ref{sec:2} presents the theoretical model of back-reflection coupling and explains the roles of single photodetector (PD) readout, balanced detection, and \(\pi\) measurement. Section~\ref{sec:3} introduces the experimental setup and its performance, followed by optical path scans and model fitting to characterize the working point dependence and establish the adjustment procedure. The suppression achieved at the identified working points is then evaluated under single-frequency and band-limited noise disturbances. Further tests show that \(\pi\) measurement retains its ability to diagnose the working point under very weak back-reflection and that the beat-amplitude difference can indicate the direction of small path offsets from the selected working point. Section~\ref{sec:4} discusses practical implementation and possible extensions of the method, and Section~\ref{sec:5} summarizes the main findings and their implications.

\section{Principle of back-reflection in transponder-type interferometry}\label{sec:2}

\begin{figure*}[t]
\centering
\includegraphics[width=\FigureOneWidth]{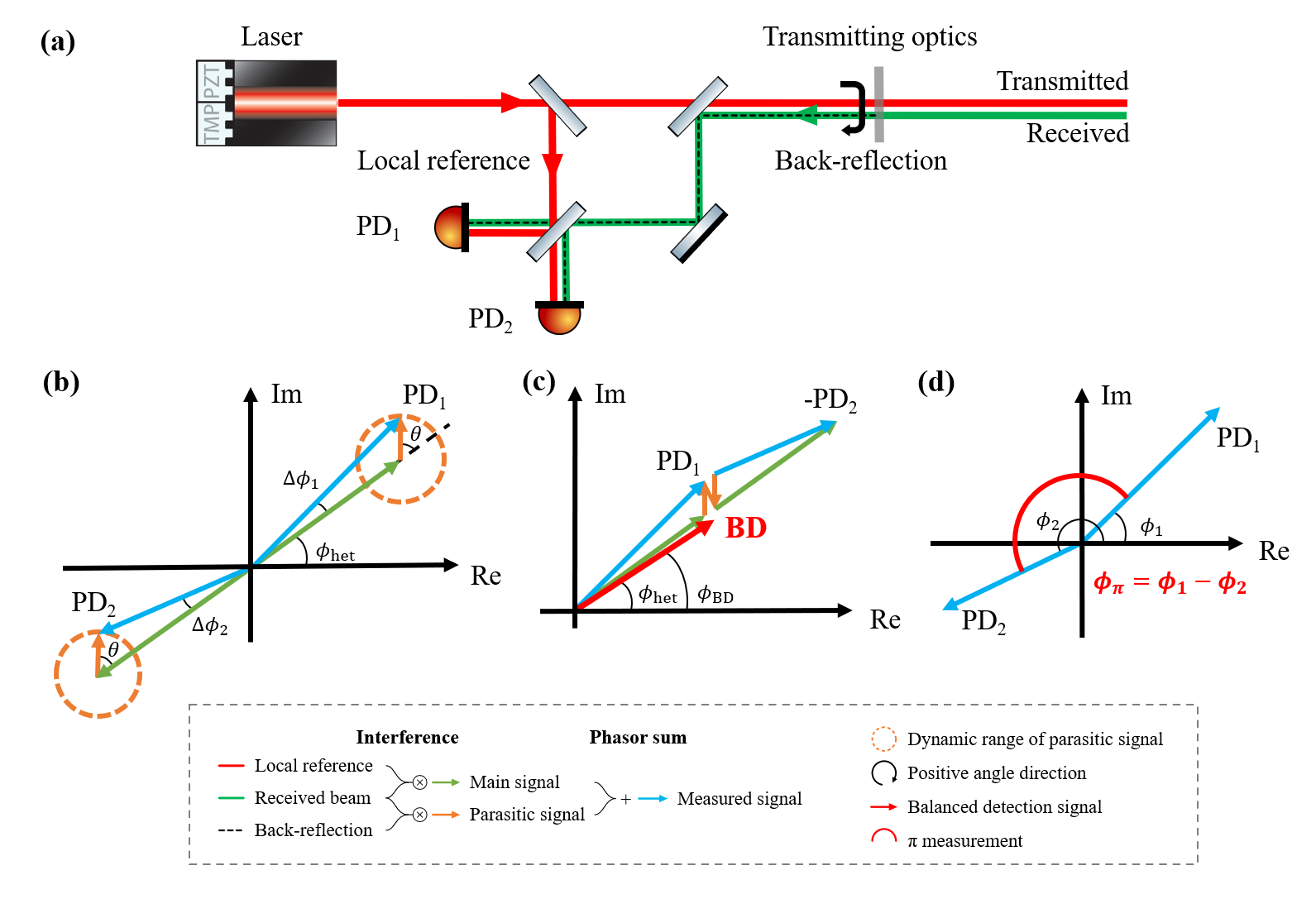}
\caption{Schematic illustration of back-reflection and parasitic interference in one satellite. (a) Optical paths, where overlapping beams are drawn separately for clarity. (b--d) Phasor diagrams of heterodyne interference with back-reflection. (b) Interference at PD\textsubscript{1} and PD\textsubscript{2}. (c) Balanced detection (BD). (d) \(\pi\) measurement. PD: photodetector.}\label{fig:1}
\end{figure*}

Fig.~\ref{fig:1}(a) illustrates back-reflection from the transmitting optics of a transponder-type interferometer, where a fraction of the transmitted beam returns toward the optical bench along the received-beam path. The local reference beam, received beam, and back-reflected beam are combined at the beam splitter (BS) and detected by the photodetectors at its symmetric and asymmetric output ports. The electric fields of local reference light, signal light, and back-reflection can be expressed as

\begin{gather}
E_{\mathrm{lo}}=A_{\mathrm{lo}}{e}^{-i(\omega _{\mathrm{lo}}t+\phi _{\mathrm{lo}})}\\
E_{\mathrm{sig}}=A_{\mathrm{sig}}{e}^{-i(\omega _{\mathrm{sig}}t+\phi _{\mathrm{sig}})}\\
E_{\mathrm{b}}=A_{\mathrm{b}}{e}^{-i(\omega _{\mathrm{lo}}t+\phi _{\mathrm{b}})}
\end{gather}

where \(A_{\mathrm{lo}}\), \(A_{\mathrm{sig}}\), and \(A_{\mathrm{b}}\) denote the field amplitudes, \(\omega _{\mathrm{lo}}\) and \(\omega _{\mathrm{sig}}\) denote the angular frequencies, and \(\phi _{\mathrm{lo}}\), \(\phi _{\mathrm{sig}}\), and \(\phi _{\mathrm{b}}\) denote the corresponding phases. The back-reflected beam has the same carrier frequency as the local reference beam because both originate from the master laser. Adopting a BS phase convention with a \(\pi /2\) phase difference between reflected and transmitted fields, the electric fields at PD\textsubscript{1} and PD\textsubscript{2} can be written as

\begin{gather}
\begin{aligned}
E_{1}&=A_{\mathrm{lo}1}{e}^{-i(\omega _{\mathrm{lo}}t+\phi _{\mathrm{lo}}+\frac{\pi }{2})}\\
&\quad +A_{\mathrm{sig}1}{e}^{-i(\omega _{\mathrm{sig}}t+\phi _{\mathrm{sig}})}\\
&\quad +A_{\mathrm{b}1}{e}^{-i(\omega _{\mathrm{lo}}t+\phi _{\mathrm{b}})}
\end{aligned}\\
\begin{aligned}
E_{2}&=A_{\mathrm{lo}2}{e}^{-i(\omega _{\mathrm{lo}}t+\phi _{\mathrm{lo}})}\\
&\quad +A_{\mathrm{sig}2}{e}^{-i(\omega _{\mathrm{sig}}t+\phi _{\mathrm{sig}}+\frac{\pi }{2})}\\
&\quad +A_{\mathrm{b}2}{e}^{-i(\omega _{\mathrm{lo}}t+\phi _{\mathrm{b}}+\frac{\pi }{2})}
\end{aligned}
\end{gather}

The detected optical power is proportional to the squared magnitude of the electric field. For PD\textsubscript{1}, it is given by

\begin{equation}
\begin{aligned}
P_{1}&=P_{\mathrm{lo}1}+P_{\mathrm{sig}1}+P_{\mathrm{b}1}\\
&\quad+2\sqrt{\eta _{\mathrm{hom}}P_{\mathrm{lo}1}P_{\mathrm{b}1}}
\cos\left(\phi _{\mathrm{lo}}+\frac{\pi}{2}-\phi _{\mathrm{b}}\right)\\
&\quad+2\sqrt{\eta _{\mathrm{het}}P_{\mathrm{lo}1}P_{\mathrm{sig}1}}\\
&\qquad{}\times\cos\left(\omega _{\mathrm{het}}t+\phi _{\mathrm{lo}}+\frac{\pi}{2}-\phi _{\mathrm{sig}}\right)\\
&\quad+2\sqrt{\eta _{\mathrm{b}}P_{\mathrm{b}1}P_{\mathrm{sig}1}}
\cos\left(\omega _{\mathrm{het}}t+\phi _{\mathrm{b}}-\phi _{\mathrm{sig}}\right)
\end{aligned}
\end{equation}

where \(\eta _{\mathrm{hom}}\), \(\eta _{\mathrm{het}}\), and \(\eta _{\mathrm{b}}\) are the corresponding interference efficiencies, and \(\omega _{\mathrm{het}}=\omega _{\mathrm{lo}}-\omega _{\mathrm{sig}}\) is the heterodyne angular frequency. The first three terms give the individual beam powers. The three cosine terms represent, respectively, the homodyne interference, the main heterodyne signal, and the parasitic heterodyne signal due to back-reflection. In heterodyne interference, normally only AC signals around the heterodyne frequency are processed, so we can express AC signals of PD\textsubscript{1} and PD\textsubscript{2}

\begin{gather}
\begin{aligned}
P_{\mathrm{AC}1}&=K_{1}\bigl[\cos  \left(\omega _{\mathrm{het}}t+\phi _{\mathrm{het}}\right)\\
&\quad+r_{1}\cos  \left(\omega _{\mathrm{het}}t+\phi _{\mathrm{het}}+\theta \right)\bigr]
\end{aligned}\\
\begin{aligned}
P_{\mathrm{AC}2}&=K_{2}\bigl[-\cos  \left(\omega _{\mathrm{het}}t+\phi _{\mathrm{het}}\right)\\
&\quad+r_{2}\cos  \left(\omega _{\mathrm{het}}t+\phi _{\mathrm{het}}+\theta \right)\bigr]
\end{aligned}
\end{gather}

where \(\phi _{\mathrm{het}}=\phi _{\mathrm{lo}}+\frac{\pi }{2}-\phi _{\mathrm{sig}}\) is the ideal interference phase unaffected by back-reflection, \(\theta =\phi _{\mathrm{b}}-\phi _{\mathrm{lo}}-\frac{\pi }{2}\) denotes the relative phase of the parasitic signal with respect to the main signal at PD\textsubscript{1}, which defines the working point. \(K_{i}=2\sqrt{\eta _{\mathrm{het}}P_{\mathrm{lo}i}P_{\mathrm{sig}i}}\) denotes the amplitude of the main interference signal, and \(r_{i}=\sqrt{\frac{\eta _{\mathrm{b}}P_{\mathrm{b}i}}{\eta _{\mathrm{het}}P_{\mathrm{lo}i}}}\) denotes the amplitude ratio of the parasitic signal to the main signal, where \textit{i} = 1, 2.

Heterodyne signals with angular frequency \(\omega _{\mathrm{het}}\) can be represented by phasors, as shown in Fig.~\ref{fig:1}(b-d). In Fig.~\ref{fig:1}(b), main signals at PD\textsubscript{1} and PD\textsubscript{2} differ in phase by \(\pi\), whereas the parasitic signals have the same phase. At each port, the measured signal is the vector sum of the main and parasitic signals, resulting in phase errors of \(\Delta \phi _{1}\) and \(\Delta \phi _{2}\). For a fixed parasitic amplitude, motion of the back-reflection source changes the relative phase \textit{\(\theta\)}, rotating the parasitic phasor and producing fluctuations in the measured phase. The phase errors, referenced to the respective main-signal phasors, are given by

\begin{gather}
\Delta \phi _{1}=\operatorname{atan2}(r_{1}\sin  \theta , 1+r_{1}\cos  \theta )\\
\Delta \phi _{2}=\operatorname{atan2}(-r_{2}\sin  \theta , 1-r_{2}\cos  \theta )
\end{gather}

where atan2(\textit{y}, \textit{x}) denotes the quadrant-aware phase angle of (\textit{x} + \textit{iy}). For \(r_{i}\)\textit{ }\(\ll\) 1, the phase errors can be approximated as

\begin{gather}
\Delta \phi _{1}\approx r_{1}\sin  \theta\\
\Delta \phi _{2}\approx -r_{2}\sin  \theta
\end{gather}

Because the main signals at the two output ports are opposite in phase whereas the parasitic signals are in phase, balanced detection can suppress the back-reflection contribution. As shown in Fig.~\ref{fig:1}(c), the signals at PD\textsubscript{1} and PD\textsubscript{2} are subtracted before phase demodulation, giving

\begin{equation}
\begin{aligned}
P_{\mathrm{BD}}&=\frac{P_{\mathrm{AC}1}-P_{\mathrm{AC}2}}{2}\\
&=\frac{K_1+K_2}{2}\Bigl[\cos(\omega_{\mathrm{het}}t+\phi_{\mathrm{het}})\\
&\quad+\frac{K_1r_1-K_2r_2}{K_1+K_2}
\cos(\omega_{\mathrm{het}}t+\phi_{\mathrm{het}}+\theta)\Bigr]
\end{aligned}
\end{equation}

The balanced detection (BD) signal has the same form as the PD\textsubscript{1} signal, with an equivalent residual amplitude ratio \(r_{\mathrm{eq}}=\frac{K_{1}r_{1}-K_{2}r_{2}}{K_{1}+K_{2}}\). The sign of \(r_{\mathrm{eq}}\) is retained to account for either direction of imbalance between the parasitic contributions at the two ports. The phase of the balanced detection signal is therefore

\begin{equation}
\begin{aligned}
\phi_{\mathrm{BD}}&=\phi_{\mathrm{het}}\\
&\quad+\operatorname{atan2}(r_{\mathrm{eq}}\sin\theta,1+r_{\mathrm{eq}}\cos\theta)
\end{aligned}
\end{equation}

Within this model, the parasitic contribution is completely canceled when \(K_{1}r_{1}=K_{2}r_{2}\), yielding \(r_{\mathrm{eq}}=0\) and \(\phi _{\mathrm{BD}}=\phi _{\mathrm{het}}\). Imperfect balance leaves a residual coupling determined by \(r_{\mathrm{eq}}\).

As shown in Fig.~\ref{fig:1}(d), \(\pi\) measurement independently demodulates the phases at PD\textsubscript{1} and PD\textsubscript{2} and then takes their difference. Referencing the ideal phase difference between the two ports to \(\pi\), the \(\pi\) measurement signal is

\begin{equation}
\begin{aligned}
\phi_{\pi}&=\pi+\Delta\phi_1-\Delta\phi_2\\
&=\pi+\operatorname{atan2}(r_1\sin\theta,1+r_1\cos\theta)\\
&\quad-\operatorname{atan2}(-r_2\sin\theta,1-r_2\cos\theta)
\end{aligned}
\end{equation}

With \textit{r}\textsubscript{1}, \textit{r}\textsubscript{2} \(\ll\)1, \(\phi _{\pi }\) can be approximated as

\begin{equation}
\phi _{\pi }\approx \pi +\left(r_{1}+r_{2}\right)\sin  \theta
\end{equation}

The back-reflection-induced phase errors therefore add in \(\pi\) measurement, whereas the common phase \(\phi _{\mathrm{het}}\) cancels. This provides an enhanced response to back-reflection while rejecting the inter-satellite displacement contribution shared by the two output ports. Moreover, suppression of common phase fluctuations gives \(\pi\) measurement a lower diagnostic noise floor than the single PD and balanced detection readouts, enabling the detection of weak back-reflection that would otherwise be obscured by background noise. Together, these properties make \(\pi\) measurement a sensitive auxiliary channel for in situ back-reflection diagnosis.

Relative phase \textit{\(\theta\)} between the main and parasitic heterodyne signals defines the back-reflection working point. In the experimental setup described in Section~\ref{sec:3.1}, reflector displacement \textit{x} changes \textit{\(\theta\)}, giving \(\theta =\theta _{0}+4\pi x/\lambda\), where \textit{\(\theta\)}\textsubscript{0} is the initial relative phase and \(\lambda\) is the laser wavelength. The local coupling coefficient is defined as \(C_{j}=d\phi _{j}/dx\), where \(\phi _{j}\) is the measured phase of each readout. Low- and high-sensitivity working points correspond to small and large values of \(|C_{j}|\), respectively, describing how strongly reflector displacement affects the measured phase. At a low-sensitivity working point, \(C_{j}\) is close to zero and the measured phase is near a local extremum. Small displacements on either side of the extremum produce phase changes of the same sign, so the phase alone cannot distinguish the two directions.

The beat amplitudes at the two output ports provide additional information about the direction of displacement. In Fig.~\ref{fig:1}(b), the beat amplitude at each port is the magnitude of the sum of the main and parasitic phasors. The beat amplitudes of PD\textsubscript{1} and PD\textsubscript{2} are therefore given

\begin{gather}
A_{1}=K_{1}\sqrt{1+2r_{1}\cos  \theta +r_{1}^{2}}\\
A_{2}=K_{2}\sqrt{1-2r_{2}\cos  \theta +r_{2}^{2}}
\end{gather}

For weak back-reflection with \textit{r}\textsubscript{1}, \textit{r}\textsubscript{2} \(\ll\)1, they can be approximated as

\begin{gather}
A_{1}\approx K_{1}(1+r_{1}\cos  \theta )\\
A_{2}\approx K_{2}(1-r_{2}\cos  \theta )
\end{gather}

The beat amplitudes thus vary in opposite directions as \textit{\(\theta\)}\textit{ }changes. Taking their difference gives

\begin{equation}
A_{1}-A_{2}\approx \left(K_{1}-K_{2}\right)+\left(K_{1}r_{1}+K_{2}r_{2}\right)\cos  \theta 
\end{equation}

Here \(K_{1}-K_{2}\) is a constant offset, while the term proportional to \(\cos  \theta\) is in quadrature with the \(\sin  \theta\) dependence of \(\pi\) measurement in Eq. (16). Near an extremum of the \(\pi\) measurement signal, the beat-amplitude difference varies approximately linearly with the back-reflection optical path. The sign of its change relative to the value at the selected working point therefore distinguishes opposite path offsets.

\section{Experiments}\label{sec:3}

\subsection{Setup and performance of the transponder-type interferometer}\label{sec:3.1}

\begin{figure*}[t]
\centering
\includegraphics[width=\FigureTwoWidth]{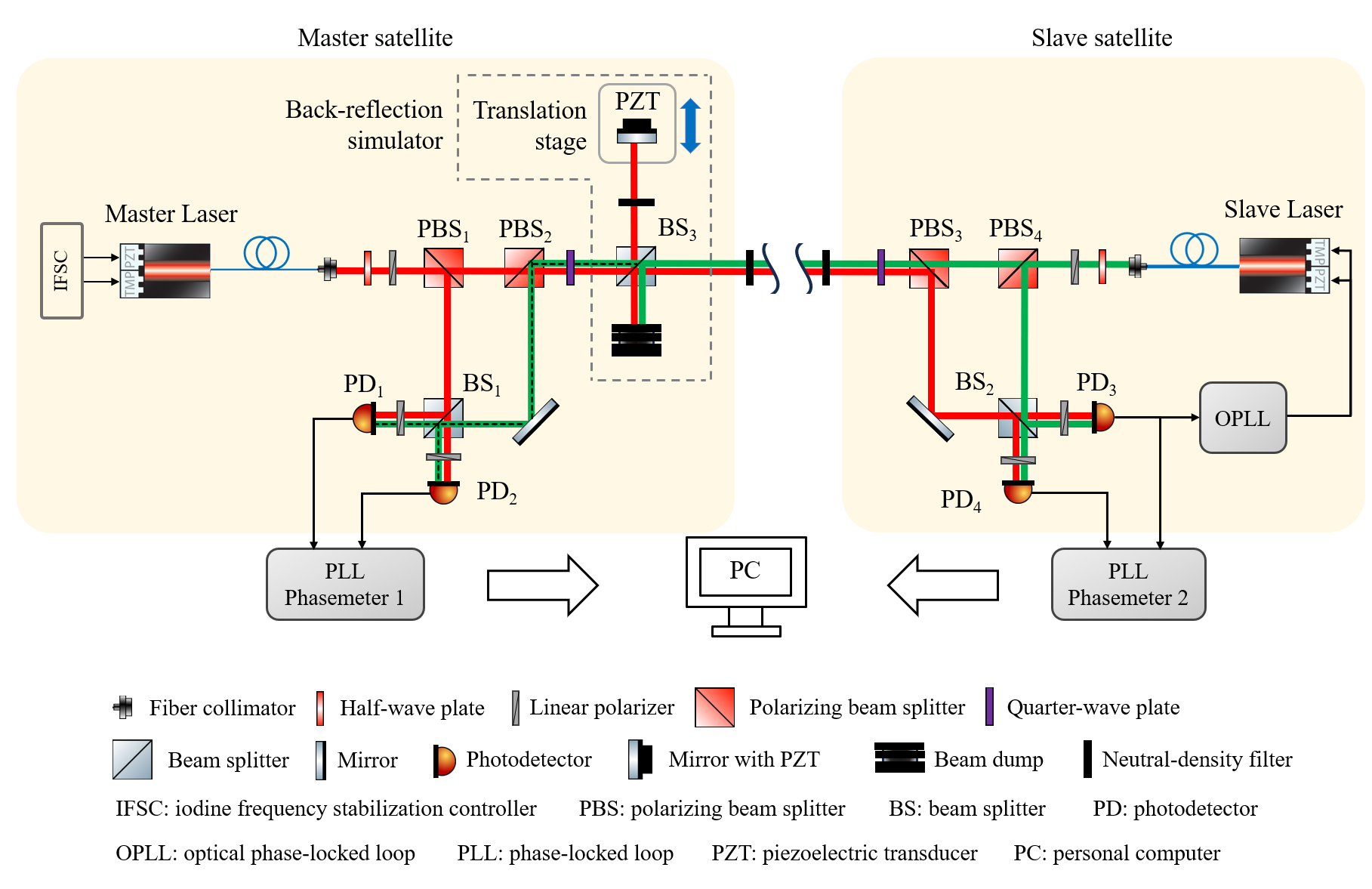}
\caption{Experimental setup of the transponder-type interferometer with coaxially aligned counter-propagating beams. Dashed outlines indicate the adjustable reflecting system that acts as a back-reflection simulator.}\label{fig:2}
\end{figure*}

To investigate parasitic back-reflection, we built a transponder-type interferometer with coaxially aligned counter-propagating beams in a laboratory environment. The experimental setup is shown in Fig.~\ref{fig:2}. The optics representing the master and slave satellites are mounted on two separate breadboards. Two neutral-density filters between the satellites simulate optical attenuation in the inter-satellite laser link. Beams from two nonplanar ring oscillator (NPRO) lasers are delivered to the respective breadboards through single-mode polarization-maintaining fibers. The master laser is frequency-stabilized using iodine absorption, while the slave laser is phase-locked to the incoming master beam through an optical phase-locked loop (OPLL) \cite{ref9,ref32}. After each fiber collimator, a half-wave plate (HWP) and a linear polarizer adjust the beam power and polarization. A~pair of polarizing beam splitters (PBSs) in each satellite separates the local reference, transmitted, and received beams. The local reference and received beams are s-polarized, whereas the transmitted beam is p-polarized. Quarter-wave plates (QWPs) convert the transmitted beams to circular polarization along the inter-satellite link and to s-polarization upon reception at the opposite satellite. The local reference and received beams are recombined at BS\textsubscript{1} and BS\textsubscript{2} to produce heterodyne signals.

An adjustable reflecting system is introduced into the master-side transmit path to simulate back-reflection. BS\textsubscript{3} directs a fraction of the transmitted beam toward a mirror mounted on a piezoelectric transducer (PZT), with the assembly installed on a translation stage. The translation stage adjusts the back-reflection optical path, while the PZT applies displacement disturbances. A~neutral-density filter in this path controls the back-reflected optical power, and a beam dump terminates the unused output of BS\textsubscript{3}. The back-reflected beam is recombined with the beam received from the slave satellite at BS\textsubscript{3} and subsequently reaches PD\textsubscript{1} and PD\textsubscript{2}. During alignment, a CCD beam profiler was used to check the spatial overlap of the received beam with both the local reference beam and the back-reflected beam. Two photodetectors at the output ports of each recombination beam splitter detect the interference signals. The phasemeters process the signals from all four PDs. The resulting phase readouts from all four PDs and beat-amplitude readouts from PD\textsubscript{1} and PD\textsubscript{2} are then recorded. The signal from PD\textsubscript{3} is also used by the OPLL to phase-lock the slave laser \cite{ref9}.

\begin{figure*}[t]
\centering
\includegraphics[width=\FigureThreeWidth]{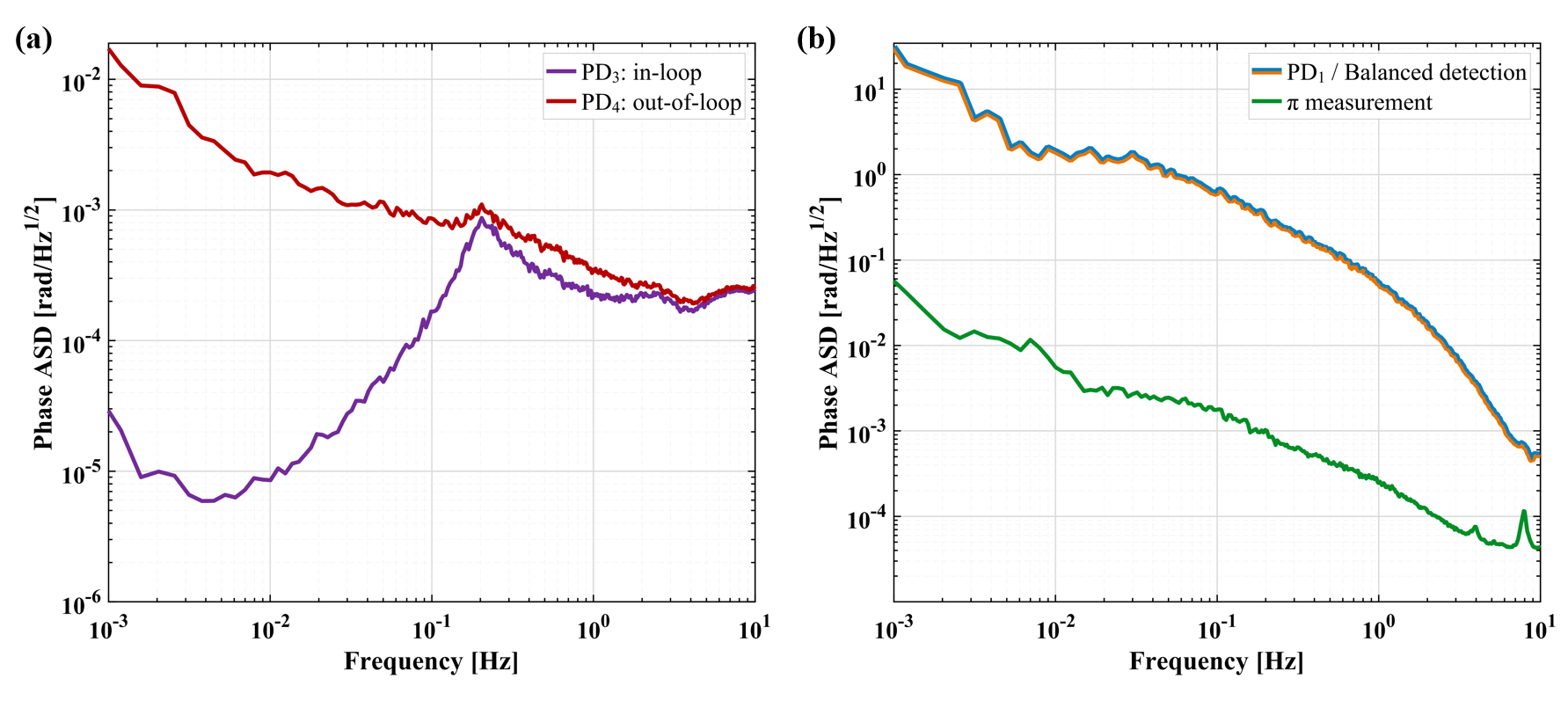}
\caption{Phase noise amplitude spectral densities (ASDs) of the transponder-type interferometer. (a) Slave-side in-loop (PD\textsubscript{3}) and out-of-loop (PD\textsubscript{4}) readouts. (b) Master-side PD\textsubscript{1}, balanced detection, and \(\pi\) measurement readouts. The nearly overlapping PD\textsubscript{1} and balanced-detection spectra are represented by a two-color line. The PD\textsubscript{2} spectrum is nearly identical to that of PD\textsubscript{1} and is omitted for clarity.}\label{fig:3}
\end{figure*}

The received optical power was determined from the photodetector output before attenuation and the calibrated power transmission ratios of two neutral-density filters. With a detector conversion gain of \NumUnit{3000}{V/W}, the PD output of \NumUnit{340}{mV} for the signal beam corresponded to an optical power of \NumUnit{110}{\UnitMicro W} before attenuation. The two filters had calibrated power transmission ratios of 1.54\(\times\)10\textsuperscript{-3} and 4.13\(\times\)10\textsuperscript{-3}, yielding a combined transmission ratio of 6.36\(\times\)10\textsuperscript{-6} and a received optical power of approximately \NumUnit{720}{pW}. This received power is on the hundreds-of-picowatts scale relevant to inter-satellite laser links in SGWIs, providing representative weak light conditions for the transponder experiment.

\begin{figure*}[t]
\centering
\includegraphics[width=\FigureFourWidth]{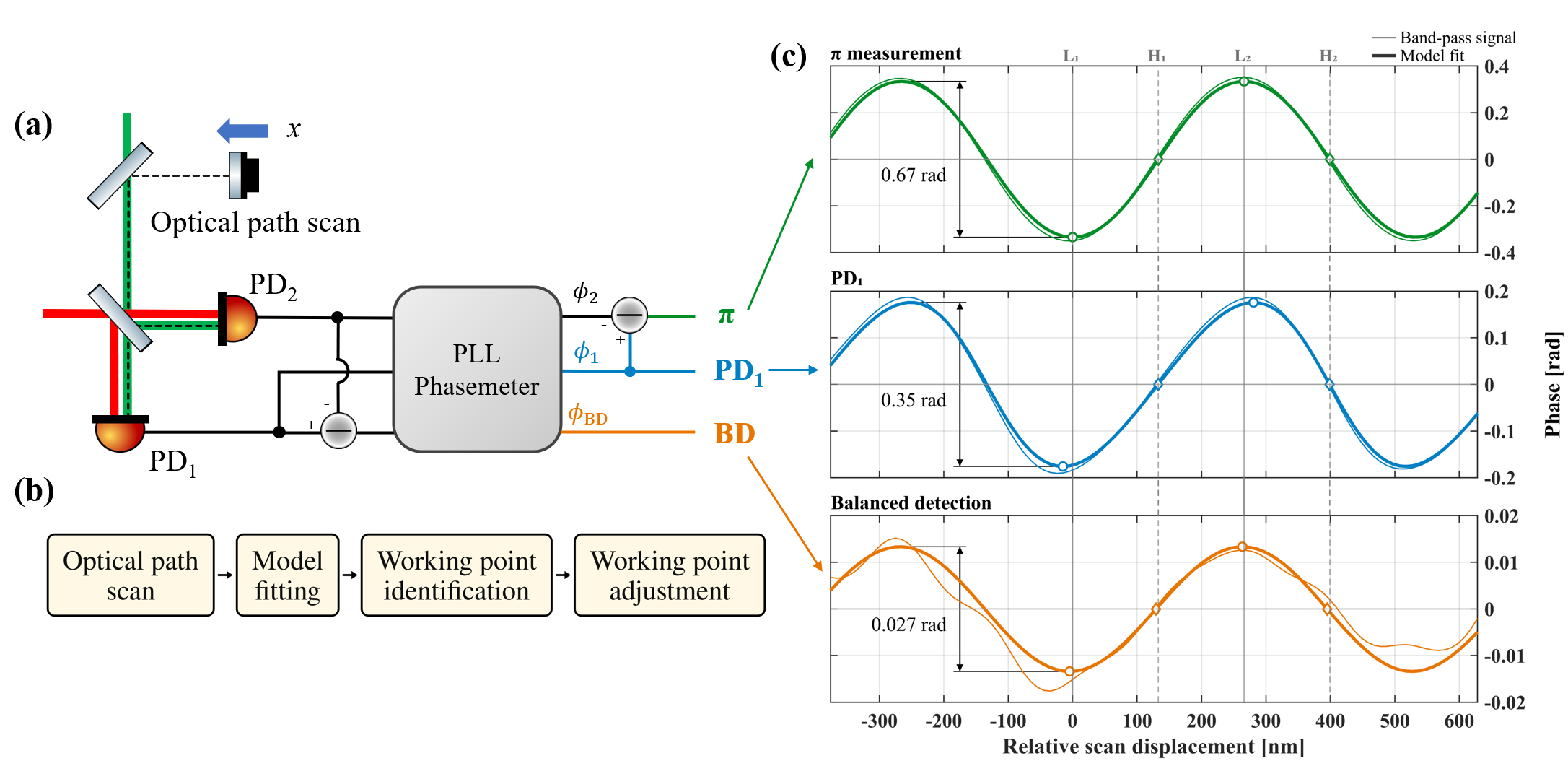}
\caption{Back-reflection characterization and \(\pi\)-assisted working point adjustment. (a) Simplified schematic of the optical path scan and three readout schemes. (b) Procedure for working point identification and adjustment. (c) Phase signals from \(\pi\) measurement, PD\textsubscript{1}, and balanced detection during an optical path scan, together with their theoretical model fits, at a received signal power of \NumUnit{720}{pW} and a back-reflection power of \NumUnit{25}{\UnitMicro W}. The annotated values indicate the fitted peak-to-peak phase variations. High-sensitivity working points (H\textsubscript{1}-point and H\textsubscript{2}-point) and low-sensitivity working points (L\textsubscript{1}-point and L\textsubscript{2}-point) are identified from \(\pi\) measurement.}\label{fig:4}
\end{figure*}

The phase noise amplitude spectral densities (ASDs) at the slave and master satellites, measured without introduced back-reflection, are shown in Fig.~\ref{fig:3}. Fig.~\ref{fig:3}(a) presents the slave-side phase-locking performance, with in-loop and out-of-loop phase ASDs of \NumUnit{2.2\times {10}^{-4}}{rad/\sqrt{Hz}} and \NumUnit{3.5\times {10}^{-4}}{rad/\sqrt{Hz}} at \NumUnit{1}{Hz}, respectively. These values correspond to equivalent optical-path noise levels of \NumUnit{38}{pm/\sqrt{Hz}} and \NumUnit{59}{pm/\sqrt{Hz}}. Fig.~\ref{fig:3}(b) compares the master-side single PD readout, balanced detection, and \(\pi\) measurement. The PD\textsubscript{1} and balanced detection ASDs nearly overlap over the measured frequency range, with both having an ASD of approximately \NumUnit{5.4\times {10}^{-2}}{rad/\sqrt{Hz}} at \NumUnit{1}{Hz}, corresponding to an equivalent optical-path noise of \NumUnit{9.2}{nm/\sqrt{Hz}}. Their noise increases substantially toward lower frequencies, where the laboratory measurements are strongly affected by environmental optical path-length fluctuations. These spectra characterize the overall noise under the current laboratory conditions. By suppressing common phase fluctuations at the two output ports, \(\pi\) measurement achieves a lower ASD of \NumUnit{2.5\times {10}^{-4}}{rad/\sqrt{Hz}} at \NumUnit{1}{Hz}. This lower diagnostic noise floor facilitates the observation of weak back-reflection-induced phase perturbations that may be obscured in the single PD and balanced detection readouts.

\subsection{Back-reflection characterization and working point adjustment}\label{sec:3.2}

To characterize the dependence of back-reflection-induced phase errors on the working point, an optical path scan was performed using the translation stage, as illustrated in Fig.~\ref{fig:4}(a). The reflecting mirror was translated at a constant velocity of \NumUnit{2}{\UnitMicro m/s}, continuously varying the relative phase between the main and parasitic heterodyne signals. The PD\textsubscript{1}, balanced detection, and \(\pi\) measurement signals were recorded simultaneously. The received signal and back-reflection powers were \NumUnit{720}{pW} and \NumUnit{25}{\UnitMicro W}, respectively. Back-reflection power was estimated from the PD output voltage using the detector conversion gain.

The procedure for \(\pi\)-assisted working point adjustment is shown in Fig.~\ref{fig:4}(b). The phase signals are first recorded during an optical path scan. The \(\pi\) measurement signal is then fitted using the model introduced in Section~\ref{sec:2} to identify high- and low-sensitivity working points. Finally, the translation stage is moved to a low-sensitivity working point for subsequent measurements. This procedure uses \(\pi\) measurement to guide working point adjustment for the single PD and balanced detection readouts.

Fig.~\ref{fig:4}(c) presents the optical path scan results. At a wavelength of \NumUnit{1064}{nm}, the stage velocity produces a periodic phase response at approximately \NumUnit{3.76}{Hz}. The recorded signals were therefore band-pass filtered from 2 to \NumUnit{15}{Hz} to suppress low-frequency drift and high-frequency noise while retaining the fundamental component and its lower harmonics. The resulting band-pass signals and theoretical model fits are shown by the thin and thick solid curves, respectively. The amplitude ratios were estimated from the band-pass signals and held fixed during the model fits. All three readouts exhibit approximately sinusoidal phase variations consistent with the theoretical models, revealing the periodic dependence of back-reflection-induced phase errors on the working point. Their fitted peak-to-peak phase variations are \NumUnit{0.67}{rad} for \(\pi\) measurement, \NumUnit{0.35}{rad} for PD\textsubscript{1}, and \NumUnit{0.027}{rad} for balanced detection. Balanced detection reduces the phase error variation by more than one order of magnitude relative to PD\textsubscript{1}, while \(\pi\) measurement exhibits a stronger response to the same back-reflection disturbance, demonstrating its sensitivity to parasitic interference.

The fitted \(\pi\) curve identifies the working points marked in Fig.~\ref{fig:4}(c). Its extrema correspond to low-sensitivity points such as the L\textsubscript{1}-point and L\textsubscript{2}-point, where the local slope approaches zero. Positions near the crossings of its mean level correspond to high-sensitivity points such as the H\textsubscript{1}-point and H\textsubscript{2}-point, where the slope magnitude is large. Thus, the same small fluctuation in the back-reflection optical path produces different phase responses at these working points, illustrating how adjustment to a low-sensitivity working point can reduce back-reflection noise. The corresponding extrema and mean-level crossings of the three readouts occur at nearly the same scan positions. This agreement supports using \(\pi\) measurement to identify a low-sensitivity working point for single PD and balanced detection readouts. The following section tests this behavior under controlled single-frequency and broadband disturbances.

\begin{figure*}[t]
\centering
\includegraphics[width=\FigureFiveWidth]{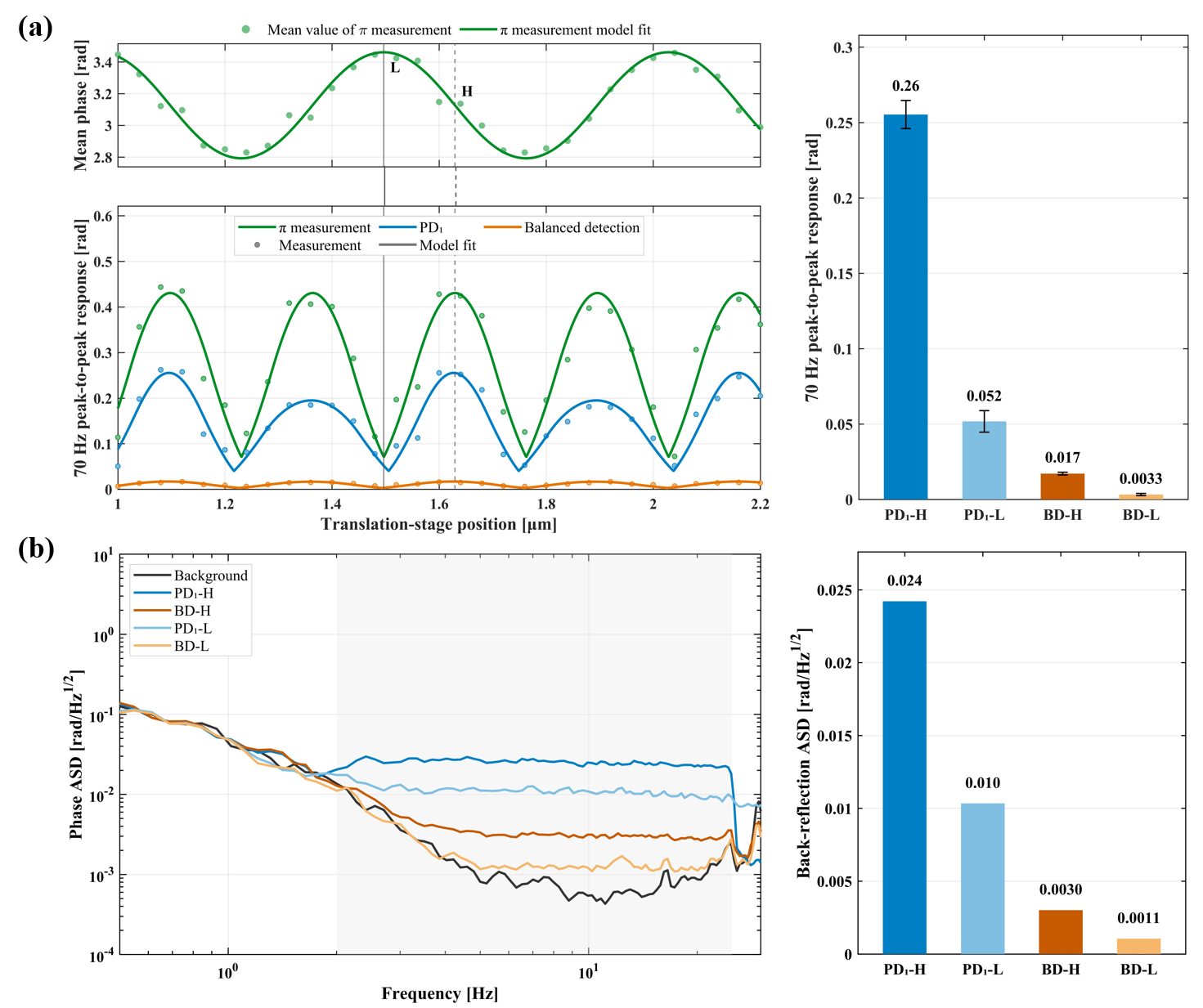}
\caption{Experimental validation of back-reflection noise suppression under single-frequency and broadband disturbances. (a) Results under a \NumUnit{70}{Hz} displacement disturbance. The upper and lower left panels show the quasi-static \(\pi\) measurement and the peak-to-peak phase responses of the three readouts as functions of translation-stage position, together with their model fits. The right panel compares the PD\textsubscript{1} and balanced detection responses at the \(\pi\)-identified H-point and L-point. (b) Results under a \NumUnit{2\text{--}25}{Hz} band-limited noise disturbance. The left panel compares the phase noise ASDs of PD\textsubscript{1} and balanced detection at the H-point and L-point, with the shaded region indicating the disturbance band. The right panel shows the band-averaged back-reflection noise ASDs after background subtraction in the power spectral density domain.}\label{fig:5}
\end{figure*}

\subsection{Experimental validation of back-reflection noise suppression}\label{sec:3.3}

To validate the suppression achieved by combining \(\pi\)-assisted working point adjustment with balanced detection, a \NumUnit{70}{Hz} displacement disturbance was applied to the back-reflection source through the PZT. The received signal and back-reflection powers were \NumUnit{720}{pW} and \NumUnit{25}{\UnitMicro W}, respectively. The translation stage was stepped through a range spanning more than two working point cycles, while the back-reflection power and disturbance amplitude were kept constant. The recorded phase signals were processed using dual-band-pass filtering around \NumUnit{70}{Hz} and its second harmonic at \NumUnit{140}{Hz} to suppress noise outside these bands. The peak-to-peak phase response was calculated as the difference between the maximum and minimum of the filtered phase signal.

The left panels of Fig.~\ref{fig:5}(a) show the quasi-static \(\pi\) measurement and the corresponding peak-to-peak phase responses of the three readouts. The response minima occur approximately at the extrema of the quasi-static \(\pi\) curve, whereas the response maxima occur near its mean-level crossings. This confirms that the \(\pi\)-identified working points correspond to different sensitivities to the applied disturbance. The right panel compares the PD\textsubscript{1} and balanced detection responses at the H-point and L-point. Error bars indicate approximate \NumUnit{95}{\%} confidence intervals propagated from the fitted model parameters. Adjusting the working point from the H-point to the L-point reduces the PD\textsubscript{1} and balanced detection responses by factors of 4.9 and 5.2, respectively. At the H-point and L-point, balanced detection reduces the response relative to PD\textsubscript{1} by factors of 15 and 16, respectively. Combining working point adjustment with balanced detection yields a maximum combined suppression factor of 77.

To evaluate the combined suppression under a broadband disturbance, a \NumUnit{2\text{--}25}{Hz} band-limited noise signal was applied to the PZT. The received signal power remained at \NumUnit{720}{pW}, while the back-reflection power was increased to approximately \NumUnit{110}{\UnitMicro W} to make the induced phase noise more clearly distinguishable from the laboratory background. The back-reflection power and applied noise signal were kept unchanged during the comparison between the H-point and L-point.

The left panel of Fig.~\ref{fig:5}(b) compares the phase noise ASDs of PD\textsubscript{1} and balanced detection at the two working points. To isolate the back-reflection contribution, the corresponding background measured without back-reflection was subtracted in the power spectral density domain, with negative differences set to zero. The resulting PSD was averaged over \NumUnit{2\text{--}25}{Hz}, and its square root was taken to obtain the band-averaged back-reflection noise ASD, as summarized in the right panel.

Balanced detection reduces the band-averaged noise ASD relative to PD\textsubscript{1} by factors of 8.0 and 9.8 at the H-point and L-point, respectively. Working point adjustment from the H-point to the L-point further reduces the PD\textsubscript{1} and balanced detection noise ASDs by factors of 2.3 and 2.9, respectively. The maximum combined suppression is obtained by comparing PD\textsubscript{1} at the H-point with balanced detection at the L-point, reducing the noise ASD from 24 to \NumUnit{1.1}{mrad/\sqrt{Hz}}, corresponding to a factor of 23. Together, the single-frequency and broadband results demonstrate that \(\pi\)-assisted working point adjustment provides additional suppression beyond balanced detection.

\begin{figure*}[t]
\centering
\includegraphics[width=\FigureSixWidth]{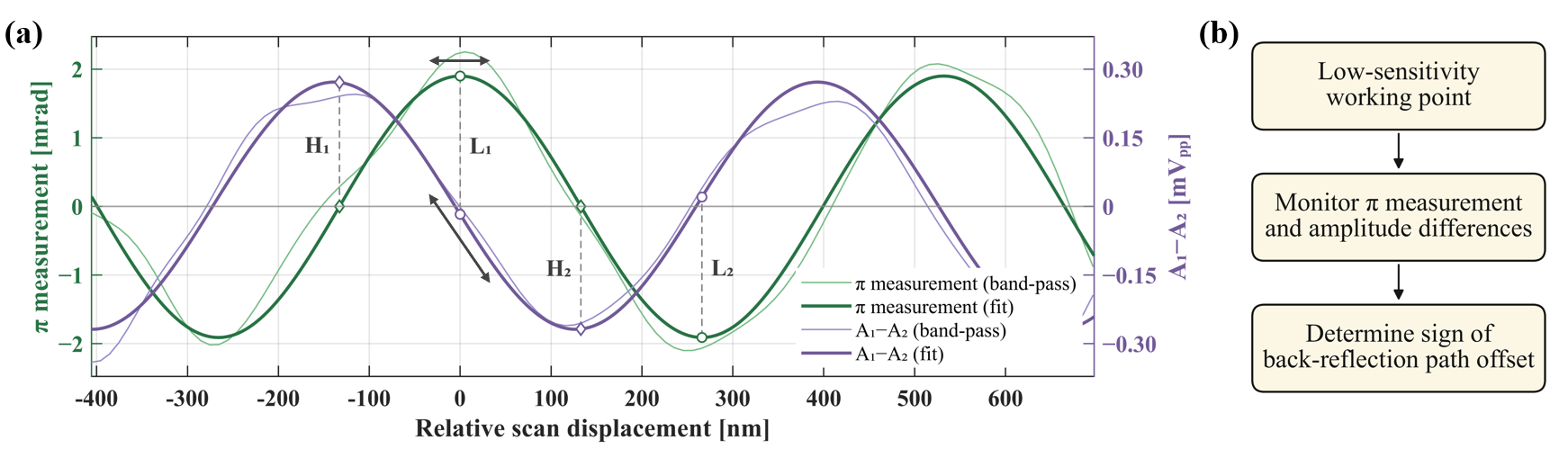}
\caption{Working point diagnosis and monitoring under weak back-reflection. (a) \(\pi\) measurement and beat-amplitude difference \(A_{1}-A_{2}\), together with their model fits, during an optical path scan at a back-reflection power of \NumUnit{1.5}{nW}. (b) Procedure for monitoring a low-sensitivity working point and determining the sign of the back-reflection path offset.}\label{fig:6}
\end{figure*}

\subsection{Working point diagnosis under extremely weak back-reflection}\label{sec:3.4}

To evaluate the diagnostic capability of \(\pi\) measurement under weak back-reflection, the optical path scan described in Section~\ref{sec:3.2} was repeated with the received signal power maintained at \NumUnit{720}{pW} and the back-reflection power reduced to \NumUnit{1.5}{nW}. The \(\pi\) measurement retained a clear periodic response. Two high-sensitivity working points and two low-sensitivity working points are marked in Fig.~\ref{fig:6}(a). Signal-to-noise ratios (SNRs) were estimated from the excess power near the fundamental scan frequency, with the noise floor estimated from adjacent frequency bands. Across ten scans, \(\pi\) measurement achieved a mean SNR of \NumUnit{26.3}{dB}, while the back-reflection response was generally difficult to resolve in the PD\textsubscript{1} and balanced detection phase readouts.

The beat-amplitude difference \(A_{1}-A_{2}\) varied approximately in quadrature with \(\pi\) measurement, as shown in Fig.~\ref{fig:6}(a). Its mean SNR reached \NumUnit{24.9}{dB}, about \NumUnit{6.1}{dB} higher than that of the PD\textsubscript{1} beat-amplitude. As illustrated by the arrows near L\textsubscript{1}-point, small path offsets to either side of the working point cause \(\pi\) measurement to decrease from its local maximum. Over the same displacement interval, \(A_{1}-A_{2}\) increases on one side and decreases on the other relative to its value at L\textsubscript{1}-point. The amplitude difference thus provides the sign information needed to distinguish the two sides of the working point.

Fig.~\ref{fig:6}(b) shows a procedure for monitoring the back-reflection optical path near a low-sensitivity working point. After the optical path is adjusted to the selected point using the procedure in Fig.~\ref{fig:4}(b), both \(\pi\) measurement and the beat-amplitude difference are monitored. For small deviations, the change in \(A_{1}-A_{2}\) from its working point value, together with the local slope in Fig.~\ref{fig:6}(a), determines the sign of the optical path offset. Using this offset as a feedback error signal could enable long-term locking at the selected working point and reduce back-reflection-induced phase noise in the transponder-type interferometer.

\begin{figure}[t]
\centering
\includegraphics[width=\FigureSevenWidth]{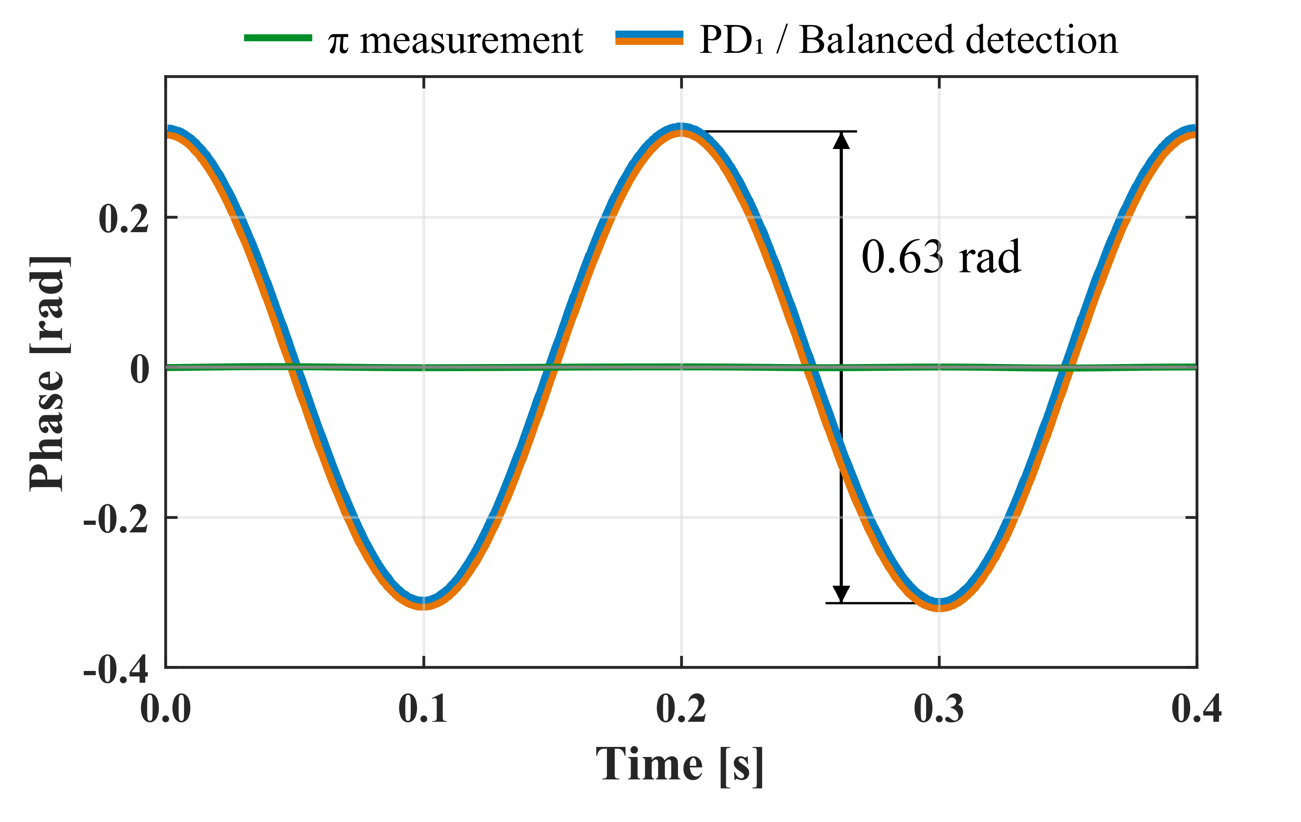}
\caption{Measured phase responses to simulated inter-satellite displacement. A \NumUnit{5}{Hz} phase modulation was applied to the OPLL in the absence of back-reflection. The traces are shown with their DC components removed. The PD\textsubscript{1} and balanced detection readouts show a peak-to-peak phase variation of approximately \NumUnit{0.63}{rad}, while \(\pi\) measurement remains close to zero.}\label{fig:7}
\end{figure}

To examine the sensitivity of \(\pi\) measurement to inter-satellite displacement, a \NumUnit{5}{Hz} phase modulation with an amplitude of \NumUnit{0.31}{rad} was applied to the OPLL without back-reflection, corresponding to a simulated inter-satellite displacement amplitude of \NumUnit{27}{nm}. As shown in Fig.~\ref{fig:7}, PD\textsubscript{1} and balanced detection exhibited nearly identical responses with a peak-to-peak response of approximately \NumUnit{0.63}{rad}, while \(\pi\) measurement remained close to zero. The coherent responses of the PD\textsubscript{1} beat amplitude and the beat-amplitude difference were also close to zero, indicating weak sensitivity to simulated inter-satellite displacement under the tested conditions. These results support the use of \(\pi\) measurement and beat-amplitude difference to monitor back-reflection while preserving the displacement signal in the balanced-detection phase readout.

\section{Discussion}\label{sec:4}

The experiments characterize the working point dependence of back-reflection coupling and demonstrate its suppression through \(\pi\)-assisted working point adjustment combined with balanced detection. Practical implementation requires consideration of environmental noise, the realization of working point adjustment, low-frequency operation, and multiple back-reflection sources.

Under weak back-reflection conditions, the contributions to the single PD and balanced detection readouts can be largely obscured by laboratory environmental noise. Difficulty in resolving these contributions does not imply that the underlying coupling is negligible. A~vacuum environment with reduced vibration and optical-path fluctuations could allow these contributions to be resolved more clearly. The ability of \(\pi\) measurement and beat-amplitude difference to retain an observable response under the same conditions demonstrates their value as sensitive auxiliary diagnostic channels.

In our experiments, back-reflection disturbances and working point adjustment were both introduced by moving the reflecting source. An alternative implementation is to adjust the optical phase of the local reference (LO) beam, thereby changing the relative phase between the main and parasitic heterodyne signals. Adjusting the LO phase changes the coupling working point but also introduces a controlled phase contribution into the measured phase. The applied phase change must therefore be calibrated and recorded, or measured by a reference interferometer, and removed from the measurement results with sufficient accuracy.

In long-baseline interferometers, low-frequency fluctuations in back-reflection optical paths are of particular concern \cite{ref13,ref24,ref33}. The test frequencies used here were selected to resolve the coupling above the laboratory environmental noise floor. As shown in Section~\ref{sec:3.4}, \(\pi\) measurement can identify a low-sensitivity working point under weak back-reflection, and the beat-amplitude difference can indicate the direction of small path offsets about that point. Together, these readouts could support feedback control to maintain the selected working point and sustain back-reflection noise suppression over long periods.

The present experiment considers a single dominant back-reflection source. In a practical interferometer containing multiple parasitic reflections, \(\pi\) measurement responds to their combined contribution and can provide information about the aggregate parasitic interference. Controlled excitation or the spectral characteristics of the response may help identify dominant contributions when necessary. Further investigation of such configurations would clarify the application of the proposed diagnosis and suppression strategy to more complex interferometers.

\section{Conclusion}\label{sec:5}

Parasitic back-reflection is an important source of phase noise in transponder-type heterodyne interferometers. Back-reflection-induced phase errors vary periodically with the working point. Working point adjustment can therefore reduce the phase response to fluctuations in the back-reflection optical path.

In the transponder-type interferometer constructed in the laboratory, \(\pi\)-assisted working point adjustment and balanced detection provided complementary noise suppression. Balanced detection reduced the parasitic contribution to the readout, while working point adjustment reduced the phase response to fluctuations in the back-reflection optical path. Their combination achieved maximum suppression factors of 77 for single-frequency phase errors and 23 for the phase-noise amplitude spectral density under band-limited disturbances, relative to single PD readout at the high-sensitivity working point. These results establish working point adjustment as an additional means of suppressing the residual noise left by imperfect balanced detection.

The value of \(\pi\) measurement extends beyond observing parasitic interference to guiding its suppression. Working points remained identifiable at a back-reflection power of \NumUnit{1.5}{nW}, even when the disturbance was largely obscured in the single PD and balanced detection phase readouts. Near a low-sensitivity working point, the beat-amplitude difference can indicate the direction of small offsets in the back-reflection optical path. Both auxiliary readouts showed little response to the simulated inter-satellite displacement, supporting distinguishing parasitic interference from the displacement being measured.

The proposed method connects back-reflection characterization, working point adjustment, and residual phase-noise suppression in transponder-type heterodyne interferometry. With improved low-frequency environmental stability and suitable feedback control, \(\pi\) measurement and the beat-amplitude difference could also support working point locking during long-term operation. The proposed strategy enables in situ stray light monitoring and suppression in precision interferometers. It could facilitate characterization and control of back-reflections from optical components, guide high-precision optical assembly, and improve the sensitivity of interferometric measurements constrained by parasitic reflections.

\begin{acknowledgments}
We gratefully acknowledge financial support from National Key Research and Development Program of China under grants 2022YFC2204001 and 2024YFC2206903; National Natural Science Foundation of China under grant 62505158.
\end{acknowledgments}

X.~Lin and X.~Xu designed the research. X.~Lin conducted the experiments, analyzed the data, and drafted the manuscript. X.~Lin and M.~Hai developed the theoretical model. X.~Lin and X.~Lyu prepared the figures. X.~Xu, Y.~Tan, Y.~Li, and H.~Wei supervised the work; X.~Xu and H.~Wei secured funding. X.~Lin, X.~Xu, X.~Lyu, P.~Tian, Y.~Feng, and H.~Wei revised the manuscript.

The authors declare no conflict of interest.

\section*{DATA AVAILABILITY}
The data are available from authors upon reasonable request.


\begin{thebibliography}{33}

\bibitem{ref1} K. Abich, A. Abramovici, B. Amparan, et al., ``In-Orbit Performance of the GRACE Follow-on Laser Ranging Interferometer,'' Phys. Rev. Lett. 123, 031101 (2019).

\bibitem{ref2} P. Auclair, D. Bacon, T. Baker, et al., ``Cosmology with the Laser Interferometer Space Antenna,'' Living Rev. Relativ. 26, 5 (2023).

\bibitem{ref3} Y. Gong, J. Luo, and B. Wang, ``Concepts and status of Chinese space gravitational wave detection projects,'' Nat. Astron. 5, 881--889 (2021).

\bibitem{ref4} The Taiji Scientific Collaboration, ``China's first step towards probing the expanding universe and the nature of gravity using a space borne gravitational wave antenna,'' Commun. Phys. 4, 34 (2021).

\bibitem{ref5} P. W. Kinman and R. M. Gagliardi, ``Doppler and range determination for deep space vehicles using active optical transponders,'' Appl. Opt. 27, 4487--4493 (1988).

\bibitem{ref6} B. Christophe, L. J. Spilker, J. D. Anderson, et al., ``OSS (Outer Solar System): a fundamental and planetary physics mission to Neptune, Triton and the Kuiper Belt,'' Exp. Astron. 34, 203--242 (2012).

\bibitem{ref7} O. Jennrich, ``LISA technology and instrumentation,'' Class. Quantum Grav. 26, 153001 (2009).

\bibitem{ref8} M. Ming, Y. Luo, Y.-R. Liang, et al., ``Ultraprecision intersatellite laser interferometry,'' Int. J. Extrem. Manuf. 2, 022003 (2020).

\bibitem{ref9} H. Mu, X. Xu, T. Le, et al., ``Transponder-type laser interferometer prototype for spaceborne gravitational wave detectors,'' Appl. Opt. 63, 1032--1038 (2024).

\bibitem{ref10} H. Mu, X. Lin, X. Xu, et al., ``Suppression of tilt-to-length coupling in transponder-type interferometer with imaging system and deliberate tilt of photodetectors,'' Opt. Lasers Eng. 186, 108753 (2025).

\bibitem{ref11} I. Khan, M. Lequime, M. Zerrad, et al., ``Detection of ultralow light power back-reflected or back-scattered by optical components using balanced low-coherence interferometry,'' Phys. Rev. Applied 16, 044055 (2021).

\bibitem{ref12} O. Gerberding and K.-S. Isleif, ``Ghost beam suppression in deep frequency modulation interferometry for compact on-axis optical heads,'' Sensors 21, 1708 (2021).

\bibitem{ref13} R. Fleddermann, C. Diekmann, F. Steier, et al., ``Sub-pm/\(\surd\)Hz non-reciprocal noise in the LISA backlink fiber,'' Class. Quantum Grav. 35, 075007 (2018).

\bibitem{ref14} K.-S. Isleif, O. Gerberding, D. Penkert, et al., ``Suppressing ghost beams: Backlink options for LISA,'' J. Phys. Conf. Ser. 840, 012016 (2017).

\bibitem{ref15} C. P. Sasso, G. Mana, and S. Mottini, ``The LISA interferometer: impact of stray light on the phase of the heterodyne signal,'' Class. Quantum Grav. 36, 075015 (2019).

\bibitem{ref16} A. Spector and G. Mueller, ``Back-reflection from a Cassegrain telescope for space-based interferometric gravitational-wave detectors,'' Class. Quantum Grav. 29, 205005 (2012).

\bibitem{ref17} B. Sang, X. Deng, W. Tao, et al., ``Stray light analysis and suppression of Taiji telescope for space gravitational wave detection based on phase noise requirement,'' Appl. Sci. 13, 2923 (2023).

\bibitem{ref18} M. Batkis, J. Berrier, K. Boyce, et al., ``Telescopes for LISA: function, current status, and path forward,'' Class. Quantum Grav. 42, 235007 (2025).

\bibitem{ref19} E. Polini and A. Chiummo, ``Stray light analysis and mitigation perspectives for next generation gravitational-wave detectors,'' Galaxies 14, 5 (2026).

\bibitem{ref20} M. Nardello, M. Lintz, and C. Buy, ``Scattering paths in test optical bench for LISA mission,'' in \textit{Frontiers in Optics / Laser Science}, OSA Technical Digest (Optica Publishing Group, 2020), paper JTu1A.6.

\bibitem{ref21} B. Canuel, E. Genin, G. Vajente, et al., ``Displacement noise from back scattering and specular reflection of input optics in advanced gravitational wave detectors,'' Opt. Express 21, 10546--10562 (2013).

\bibitem{ref22} M. Andrés-Carcasona, A. Macquet, M. Martínez, et al., ``Study of scattered light in the main arms of the Einstein Telescope gravitational wave detector,'' Phys. Rev. D 108, 102001 (2023).

\bibitem{ref23} S. Zeidler, T. Akutsu, Y. Torii, et al., ``Measuring scattering light distributions on high-absorptive surfaces for stray-light reduction in gravitational-wave detectors,'' Opt. Express 27, 16890--16910 (2019).

\bibitem{ref24} P. Wang, J. Sanjuan, M. Mehmet, and F. Guzman, ``Parasitic interference in heterodyne interferometers: Modeling, characterization, and mitigation,'' \textit{Phys. Rev. Applied} 26, 054064 (2026).

\bibitem{ref25} K.-S. Isleif, L. Bischof, S. Ast, et al., ``Towards the LISA backlink: experiment design for comparing optical phase reference distribution systems,'' Class. Quantum Grav. 35, 085009 (2018).

\bibitem{ref26} T. S. Schwarze, G. Fernández Barranco, D. Penkert, et al., ``Picometer-stable hexagonal optical bench to verify LISA phase extraction linearity and precision,'' Phys. Rev. Lett. 122, 081104 (2019).

\bibitem{ref27} J. Wang, K. Qi, H. Liu, et al., ``An integrated fiber phase modulation device for optical path noise suppression,'' Class. Quantum Grav. 43, 015001 (2026).

\bibitem{ref28} G. Hechenblaikner, V. Wand, M. Kersten, et al., ``Digital laser frequency control and phase-stabilization loops in a high precision space-borne metrology system,'' IEEE J. Quantum Electron. 47, 651--660 (2011).

\bibitem{ref29} B. Cao, F.-L. Jia, M.-L. Yang, et al., ``Suppression of frequency-mixing effect for pm-level heterodyne interferometers based on `zero coupling' optical path length control,'' Opt. Lett. 49, 3300--3303 (2024).

\bibitem{ref30} A. E. Rosenbluth and N. Bobroff, ``Optical sources of non-linearity in heterodyne interferometers,'' Precis. Eng. 12, 7--11 (1990).

\bibitem{ref31} S. Yokoyama, Y. Hori, T. Yokoyama, and A. Hirai, ``A heterodyne interferometer constructed in an integrated optics and its metrological evaluation of a picometre-order periodic error,'' Precis. Eng. 54, 206--211 (2018).

\bibitem{ref32} X. Xu, J. Liu, H. Mu, Y. Li, and Y. Tan, Six-degrees-of-freedom test mass readout via optical phase-locking heterodyne interferometry, IEEE Trans. Instrum. Meas. 73, 7007407 (2024).

\bibitem{ref33} K. Jersey, H. Hollis, H.-Y. Chia, et al., ``Picometer sensitive prototype of the optical truss interferometer for LISA,'' Class. Quantum Grav. 42, 035008 (2025).

\end{thebibliography}
\end{document}